\documentclass[twocolumn]{aastex631}
\usepackage{hyperref}
\usepackage{changepage}
\hypersetup{
    colorlinks=true,
    linkcolor=blue,
    filecolor=magenta,      
    urlcolor=cyan,
}

\usepackage{xspace}
\usepackage{float}
\usepackage{ucs}
\usepackage[utf8x]{inputenc}
\usepackage[version=4]{mhchem}
\usepackage{csvsimple}

\usepackage[caption=false]{subfig}
\usepackage[T1]{fontenc}

\newcommand{\kms}{\ensuremath{\mathrm{km}\,\mathrm{s}^{-1}}\xspace}

\newcommand{\Rsun}{\ensuremath{R_{\odot}}\xspace }
\newcommand{\Msun}{\ensuremath{M_{\odot}}\xspace}

\newcommand{\tess}{\textit{TESS} }

\begin{document}

\nolinenumbers
\title{A Small Brown Dwarf in an Aligned Orbit around a Young, Fully-Convective M Star }

\correspondingauthor{Madison Brady}
\email{mtbrady@uchicago.edu}

\author[0000-0003-2404-2427]{Madison Brady}
\affiliation{Department of Astronomy \& Astrophysics, University of Chicago, Chicago, IL 60637, USA}

\author[0000-0003-4733-6532]{Jacob L.\ Bean}
\affiliation{Department of Astronomy \& Astrophysics, University of Chicago, Chicago, IL 60637, USA}

\author[0000-0001-7409-5688]{Guðmundur Stefánsson} 
\affil{Anton Pannekoek Institute for Astronomy, University of Amsterdam, Science Park 904, 1098 XH Amsterdam, The Netherlands} 

\author[0009-0003-1142-292X]{Nina Brown}
\affiliation{Department of Astronomy \& Astrophysics, University of Chicago, Chicago, IL 60637, USA}

\author[0000-0003-4526-3747]{Andreas Seifahrt}
\affiliation{Gemini Observatory/NSF NOIRLab, 670 N. A'ohoku Place, Hilo, HI 96720, USA}

\author[0000-0003-4508-2436]{Ritvik Basant}
\affiliation{Department of Astronomy \& Astrophysics, University of Chicago, Chicago, IL 60637, USA}

\author[0009-0005-1486-8374]{Tanya Das}
\affiliation{Department of Astronomy \& Astrophysics, University of Chicago, Chicago, IL 60637, USA}

\author[0000-0002-4671-2957]{Rafael Luque}
\affiliation{Department of Astronomy \& Astrophysics, University of Chicago, Chicago, IL 60637, USA}
\affiliation{NHFP Sagan Fellow}

\author[0000-0002-4410-4712]{Julian St{\"u}rmer}
\affiliation{Landessternwarte, Zentrum f{\"u}r Astronomie der Universität Heidelberg, K{\"o}nigstuhl 12, D-69117 Heidelberg, Germany}

\begin{abstract}
A star's spin-orbit angle can give us insight into a system's formation and dynamical history.  In this paper, we use MAROON-X observations of the Rossiter-McLaughlin (RM) effect to measure the projected obliquity of the LP\,261-75 (also known as TOI-1779) system, focusing on the fully-convective M dwarf LP\,261-75A and the transiting brown dwarf LP\,261-75C.  This is the first obliquity constraint of a brown dwarf orbiting an M dwarf and the seventh obliquity constraint of a brown dwarf overall.  We measure a projected obliquity of $5^{+11}_{-10}$\,degrees and a true obliquity of $14^{+8}_{-7}$\,degrees for the system, meaning that the system is well-aligned and that  the star is rotating very nearly edge-on, with an inclination of $90^o\,\pm\,11^o$.  The system thus follows along with the trends observed in transiting brown dwarfs around hotter stars, which typically have low obliquities.  The tendency for brown dwarfs to be aligned may point to some enhanced obliquity damping in brown dwarf systems, but there is also a possibility that the LP\,261-75 system was simply formed aligned.  In addition, we note that the brown dwarf's radius ($R_C\,=\,0.9$\,R$_J$) is not consistent with the youth of the system or radius trends observed in other brown dwarfs, indicating that LP\,261-75C may have an unusual formation history.

\end{abstract}


\section{Introduction}
\label{sec:intro}

A system's obliquity ($\psi$) is the angle between a star's rotation axis and the angular momentum of its companions' orbits.  The obliquity of a planetary system can give us insight about a system's formation and dynamical history.  Alignment (zero obliquity) makes sense assuming a planet and its stars form out of the same rotating gas cloud, and implies a dynamically inactive system \citep[or, alternatively, alignment due to tidal interactions between the star and its companions- see, e.g.,][and the references therein]{Albrecht2022}.  Misalignment may be a sign of past dynamical interactions, but some studies have shown that some systems may form misaligned, due to magnetic disk-star interactions \citep{Lai11} or chaotic accretion processes during formation \citep{Takaishi2020}, though chaotic accretion cannot account for $\psi\,>\,20^o$.

Measuring $\psi$ directly is difficult.  However, we can measure a star's \textit{projected} obliquity ($\lambda$) using the Rossiter-McLaughlin (RM) effect \citep{Rossiter1924, McLaughlin1924}, which utilizes high-precision radial velocity (RV) measurements of the host star in order to track the transiting companion's passage over the surface of the star.   As the companion travels over the rotating star's surface, it will eclipse blueshifted and redshifted portions of the star, making the stellar surface appear to be ``redder'' and ``bluer'' respectively.  This will cause a RV anomaly whose precise shape depends on the orientation of the companion's orbit.

The RM effect has been used to great success to study the orbits of many exoplanet systems \citep[e.g.,][]{Queloz2000, Winn2005, Winn2006}.  Several trends have emerged from the ensemble of well-characterized planet orbits to-date.  It appears that planets orbiting stars with $T_\mathrm{eff}\,>\,6250$\,K tend to have misaligned orbits, while cooler stars tend to have more aligned orbits \citep{Winn2010b}.  The location of this break point is comparable to the Kraft break \citep{Kraft1967}, above which stars lack convective envelopes.  This seems to imply that the characteristics of a star's convective envelope could influence the realigning timescale (or primordial alignment) of its companions, with highly convective stars having more aligned companions \citep[see, e.g.,][]{Albrecht2022}.

However, it is difficult to extrapolate this trend to fully-convective stars \citep[which are stars with masses below about 0.35\,\Msun, see][]{Chabrier97}, as there are relatively few obliquity measurements of M dwarfs.  Currently, there are only seven systems around M dwarfs with measured obliquities: AU Mic \citep{Addison2021}, TRAPPIST-1 \citep{Hirano2020, Brady23}, GJ 436 \citep{Bourrier2022}, GJ 3470 \citep{Stefansson2022}, K2-25 \citep{Stefansson2020}, K2-33 \citep{Hirano2024}, and TOI 4201 \citep{Gan2024}.  The small number of obliquities is due to several factors.  Firstly, M dwarfs tend to be dim, making it difficult to perform the high-precision spectrograph measurements necessary to measure a planet's obliquity.  Additionally, the RM effect signal is directly proportional to the star's rotation velocity, and there are relatively few known M dwarfs with both high rotational velocities and known transiting planets.  This is exacerbated by the fact that rapidly-rotating M dwarfs tend to be highly active, which can obscure transiting planet signals in their photometry. 

Of the well-studied M dwarf systems, two (AU Mic and K2-33) have ages below 30\,Myr \citep{Mamajek2014, Mann16}.  Both stars have strong activity signals in their RVs that complicate the measurements of their obliquities (and lower their precision), but both are consistent with being well-aligned.  If we consider the obliquity damping timescale for stars with convective envelopes (utilizing equilibrium tide theory) from \cite{Albrecht2012}, we find that both systems are unlikely to have undergone significant realignment over their lifetimes. It is thus reasonable to assume these obliquities are primordial. This seems to imply that M dwarf systems may tend to have low primordial obliquities, but additional (and higher-precision) measurements of young M dwarf planet obliquities are necessary to support this hypothesis.

Turning our attention to the remaining older M dwarfs, we expect that, due to the convective nature of their host stars, their planets would have low obliquities.  However, two of the five systems (GJ 436 and GJ 3470) have misaligned, near-polar orbits.  This is especially interesting given their similar masses (within a factor of two) to one of the well-aligned M dwarfs, K2-25.  The obliquity damping timescales discussed in the previous paragraph indicate that none of these three systems have had time to damp their stars' obliquities, so it is unexpected that one is aligned while the other two are not.  While this could be the result of primordial obliquity differences, it is also notable that the two misaligned planets orbit stars with $M\,>\,0.35\,$\Msun (and are thus unlikely to be fully convective), while K2-25's host is likely fully convective, with a mass of 0.26\,M$_\odot$ \citep{Stefansson2020}.  This seems to point to some change in obliquity damping processes at the fully-convective boundary.  However, given the small sample size, we need to increase the number of M dwarfs with well-characterized obliquities in order to explore this trend.

In this paper, we measure the obliquity of the LP\,261-75 system.  LP\,261-75A is a small, nearby mid-M dwarf.  It has a mass of around 0.3\,$M_\odot$ \citep{Irwin2018, Paegert21}, meaning that it is likely to be fully convective \citep{Chabrier97}. LP\,261-75 is fairly unique compared to other M dwarfs with RM measurements.  Its membership in the AB Doradus Moving Group \citep{Sun2022} means that it is likely around 100\,Myr old, making it intermediate in age between K2-33 and the older M dwarfs with measured obliquities.

Unlike the other young M dwarfs, however, LP\,261-75A has an extremely large transiting companion- the brown dwarf LP\,261-75C is about 30\% the radius of the primary.  As the strength of the RM signal is proportional to the ratio of the radii of the companion to the host squared, it provides us with a uniquely large RM signal to study.  LP\,261-75C also has a $<\,2$d orbital period that places it well within the brown dwarf desert, a well-known under-density of brown dwarfs at orbital periods $<\,100$\,days  \citep{Grether2006}, which may be due to their different formation mechanisms compared to planets.  There are currently only RM obliquity measurements for six other brown dwarf systems: CoRoT-3b \citep{Triaud2009}, KELT-1b \citep{Siverd2012}, WASP-30b \citep{Triaud2013}, HATS-70b \citep{Zhou2019}, GPX-1b \citep{Giacalone2024}, and TOI-2533b \citep{Ferreira2024}. Additionally, there are no obliquity measurements for brown dwarfs around M dwarfs.

Despite their different formation mechanisms, brown dwarfs may still have high obliquities.  \cite{Jennings2021} showed that the turbulence and collisions between fragments in a disk can create brown dwarfs with obliquities up to 90$^o$.  While \cite{Marcussen2022} found that the majority of close double stars are aligned, their studied sample primarily consisted of stars with much earlier spectral types than LP\,261-75A, making it difficult to generalize their results.  LP\,261-75C's large radius, short orbital period and rapidly-rotating host star thus give us the unique opportunity to perform a very high-precision RM measurement of a star in a young system, bridging the gap between the uncertain obliquities of the young M dwarfs and the precise obliquities of the older M dwarfs.  It also provides us with the chance to study a young brown-dwarf desert object in a system with a well-characterized age.  This is the first measurement of a brown dwarf obliquity around a cool host star. 

In Section~\ref{sec:system}, we will summarize the properties of the LP\,261-75 system.  In Section~\ref{sec:obs}, we describe the MAROON-X RVs that we have collected on the LP\,261-75A, as well as all other observations available on the system.  Section~\ref{sec:analysis} details our analyses on the photometry, spectra, and RVs of the system.  Section~\ref{sec:discussion} touches on the implications our our results, and Section~\ref{sec:conclusions} concludes.

\section{The LP 261-75 System}
\label{sec:system}

LP\,261-75 is a system consisting of a 0.3\,$M_\odot$ mid-M dwarf (LP\,261-75A) and two brown dwarfs.  The first brown dwarf, LP\,261-75B, is a long-period companion with an orbital separation of approximately 450 AU and a mass of around 20\,$M_\mathrm{J}$ \citep{Reid2006}.  The second, LP\,261-75C, transits LP\,261-75A and has a radius of $\approx$0.9\,$R_\mathrm{J}$, a mass of $\approx$70\,$M_\mathrm{J}$, and an orbital period of 1.88d \citep{Irwin2018}.

The system is associated with the AB Doradus Moving Group \citep{Sun2022}, which has an age of 133$^{+15}_{-20}$\,Myr \citep{Gagne2018}.  The M dwarf is thus young and expected to be active.  Its activity is obvious in its \textit{TESS} photometry, which shows modulations consistent with a short rotation period.  \cite{CantoMartins20} found a rotation period of $1.105\,\pm\,0.027$\,days, but the additional Sector 48 data studied by \cite{Bowler2023} showed that the 1.1\,days signal was an alias and the star actually has a rotation period of around 2.2\,days, with the half-period signal the result of starspots on opposite sides of the star.  We will thus quote a rotation period of $2.214\,\pm\,0.040$\,days for the star, doubling the nominal period found in \cite{Bowler2023}.

The rotation signal likely comes from LP\,261-75A despite the blending of the three sources in the LP\,261-75 system in the \tess aperture given its relative luminosity in the \tess bandpass.  LP\,261-75A's rotation period is consistent with the expectations for a mid-M dwarf at the age of the AB Doradus Moving Group according to the relations from \cite{Engle2023}.

The stellar parameters of LP\,261-75A are provided in Table~\ref{tab:host}, and parameters relevant to LP\,261-75C are in Table~\ref{tab:bd}.

\begin{table}[]
    \centering
    \begin{tabular}{|l|cc|cc|cc|}
    \hline
    \textbf{Property}      & \textbf{Value} & \textbf{Reference} \\
    \hline
    RA (J2000)             & 06h 51m 04.58s & a \\
    Declination (J2000)    & +35$^o$ 58' 09.46" & a \\
    Distance (pc)          & 33.995 $\pm$ 0.028 & a \\
    Spectral Type          & M4.5V & b \\
    $K$ Mag                & 9.690 & c\\
    M$_A$ (\Msun)          & 0.300 $\pm$ 0.015 & d\\
    R$_A$ (\Rsun)          & 0.308 $\pm$ 0.005 & This work \\
    T$_\mathrm{eff}$ (K)   & 3138 $\pm$ 157 &  e \\
    P$_{\mathrm{rot}}$ (d) & 2.214 $\pm$ 0.040 &  f \\
    $v_{eq}$\,sin$i_\star$ (km\,s$^{-1}$)& 7.0\,$\pm$\,0.1  & This work \\
    \hline
    \end{tabular}
    \caption{Properties of LP\,261-75A, the host star.  References: a) \cite{GaiaEDR3}, b) \cite{Reid2006}, c) \cite{2MASS}, d) \cite{Irwin2018}, e) TIC v8.2 \citep{Paegert21}, f) \cite{Bowler2023}.}
    \label{tab:host}
\end{table}

\begin{table}[]
    \centering
    \begin{tabular}{|l|c|}
    \hline
    \textbf{Property}      & \textbf{Value}  \\
    \hline
    P$_{orb}$              & $1.88172235^{+0.00000009}_{-0.00000010}$ \\
    $e$ (95\% upper limit)                   &  $<0.007$ \\   
    $M_C$ ($M_J$)          & 67.4 $\pm$ 2.1 \\
    $R_C$ ($R_J$)          & $0.903 ^{+0.015}_{-0.014}$ \\
    $\lambda$ (degrees)    & $4.8^{+11.3}_{-10.2}$ \\
    $\psi$ (degrees)       & $14.0 ^{+7.8}_{-6.7}$ \\
    \hline
    \end{tabular}
    \caption{Properties of LP\,261-75C, the transiting brown dwarf.  All values are from this work except for the eccentricity, which is from \cite{Irwin2018}.}
    \label{tab:bd}
\end{table}

\section{Observations}
\label{sec:obs}

\subsection{Photometry}
\label{ssec:photometry}

The LP\,261-75 system was observed by the Transiting Exoplanet Survey Satellite \citep[\textit{TESS}][]{Ricker15}. \tess is an all-sky photometric survey that surveys large sections of the sky for 27\,day sectors.  LP\,261-75 was observed in two sectors, Sector 21 and Sector 48, with data from January-February 2021 and January-February 2023.  The data are available at a 120\,second cadence from the \tess SPOC pipeline \citep{Jenkins16}. 

\begin{figure}
    \centering
    \includegraphics[width=0.9\linewidth]{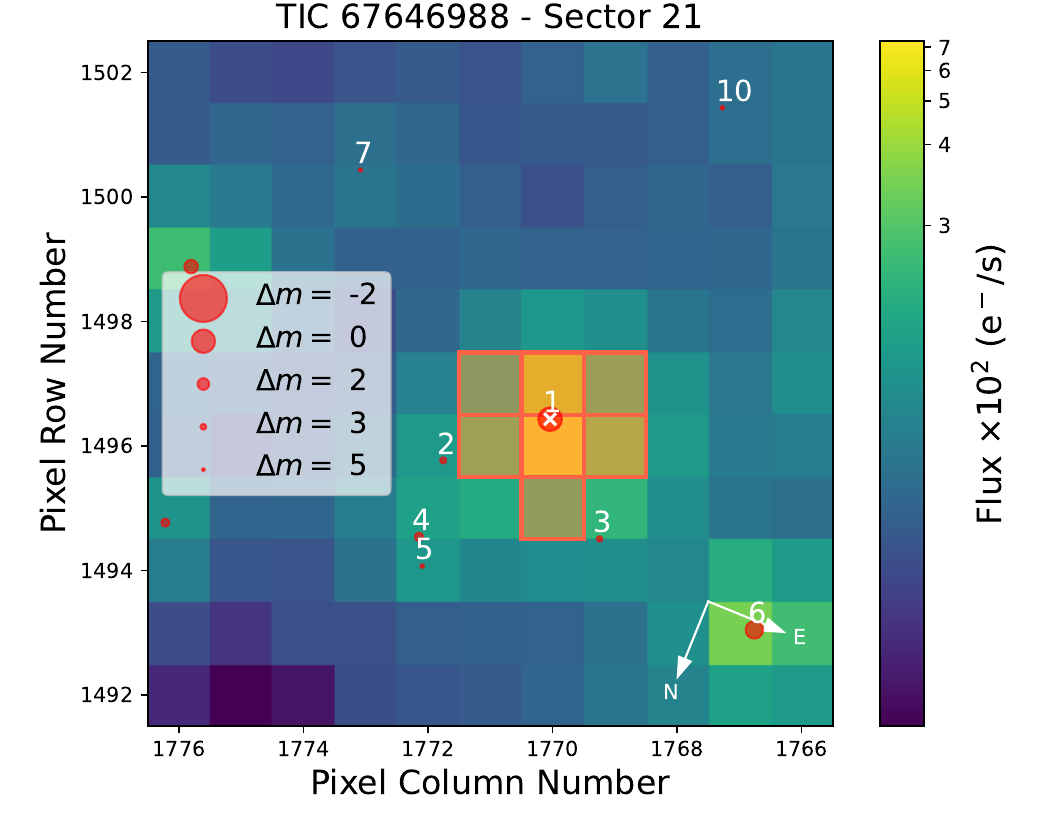}
    \caption{The field around LP\,261-75 in \tess sector 21, with the pipeline aperture outlined in red.  Red circles indicate other stars known from Gaia DR3, with their size indicating their flux relative to LP\,261-75.  The results from sector 48 are similar and thus not included.  This figure was generated using \texttt{tpfplotter}.}
    \label{fig:TOI1779_tpf}
\end{figure}

We used \texttt{tpfplotter} \citep{Aller20} to investigate the crowding of stars around LP\,261-75, as spectral contamination could influence the derived radius of the transiting LP\,261-75C.  Figure~\ref{fig:TOI1779_tpf} shows the crowding of known \textit{Gaia} sources from Gaia DR3 \cite{GaiaDR3} with $\Delta_G\,<\,6$\,mag.  It is obvious that there are no targets that fall within the pipeline aperture that are both near LP\,261-75 and bright enough to cause significant $>\,1\%$ contamination.

We note that LP\,261-75B and LP\,261-75C are not Gaia sources, and are thus not accounted for in this estimate. LP\,261-75B is around 12\,arcseconds away from LP\,261-75A \citep{Reid2006} and LP\,261-75C's short orbital period places it less than 1\,arcsecond away from the primary.  The two brown dwarfs thus fall on the same pixel as the M dwarf, making them potential contaminators.  However, they are unlikely to be a source of significant contamination.  LP\,261-75B is 6.65\,magnitudes dimmer than LP\,261-75A in the \textit{J} band \citep{2MASS}.  Given the relative temperatures of the two objects, this difference is likely to be even more stark in \textit{TESS} magnitudes.  If LP\,261-75C is similarly dim, it is also unlikely to contribute much flux in the \textit{T} band.

For the purposes of our analysis, we used the 120\,second cadence data for both sectors from the SPOC pipeline, specifically focusing on the Presearch Data Conditioning Single Aperture Photometry (PDCSAP) data.  The PDCSAP pipeline accounts for the effects of crowding and instrumental systematics.  The PDCSAP pipeline is described in more detail in \cite{Stumpe2012}, \cite{Smith2012}, and \cite{Stumpe2014}.  We downloaded the PDCSAP photometry using the \texttt{lightkurve} \citep{lightkurve}.

As discussed in \cite{Irwin2018}, there is also data from the MEarth photometric survey \citep[see][for details on target selection]{Nutzman2008} for the LP\,261-75 system.  The system was initially detected by MEarth's real-time trigger, with additional follow-up observations to measure the orbital parameters.  The MEarth observations are not evenly sampled, and instead have a cadence of around once a minute during transits and eclipses of LP\,261-75C with less data out-of-transit, limiting their usefulness for characterizing stellar rotation signals.  They encompass around 127\,days worth of data.  We used the data included in \cite{Irwin2018} for later analyses.

\subsection{Radial Velocities}
\label{ssec:rvs}

We observed LP\,261-75A in 31 exposures on 17 April, 2024 between 05:30---09:00 UTC with the MAROON-X instrument.  MAROON-X is a Extreme Precision Radial Velocity (EPRV) spectrograph installed on Gemini-North \citep{Seifahrt18, Seifahrt22}.  Given its red-optical wavelength coverage and mounting on an eight-meter telescope, it is ideally suited for studies of dim, red stars.

MAROON-X has two channels with separate CCD chips, a ``red'' channel from 650---900\,nm and a ``blue'' channel from 500---670\,nm.  As the two channels are independent and encompass different wavelength ranges, they can be treated as separate instruments for the purposes of RV analysis.  In our observations, we exposed both channels simultaneously (with 300s exposures in the red channel and 340s exposures in the blue channel), meaning that our final dataset consists of 62 RVs.  

The MAROON-X data was reduced using the standard MAROON-X pipeline (which is a custom \texttt{Python3} pipeline) and we calculated the RVs with a modified version of \texttt{serval} \citep{Zechmeister20} that works with MAROON-X data.  We also corrected the times recorded by MAROON-X to the times of the solar system barycenter using \texttt{astropy} \citep{Astropy1, Astropy2, Astropy3}.  \texttt{serval} calculates an individual spectrum's RV by comparing it to a spectral template formed by co-adding all of the spectra together.  Given the large expected RV signal of the brown dwarf (on the order of several km\,s$^{-1}$), merely co-adding the individual spectra to form the template would result in a dramatic smearing of the individual spectral lines.  Thus, we had to iterate upon our template several times, shifting the spectra by the calculated RVs in each step before co-adding them to form the template. We repeated this process until our results converged.  We found that \texttt{serval} had a tendency to overfit the red-channel spectra when spline fitting the data in order to perform the RV shifting and coadding necessary for the template creation.  This is likely due to a combination of broad spectral lines, extreme RV shifts, and a low signal-to-noise ratio (SNR).  We thus had to set an upper limit of 500 on the number of knots used to form the spline when fitting the red-channel data.  As the brown dwarf is very large compared to the host star, we expected that it would cause perturbations in the line shapes during the transit.  To account for this, we only used the out-of-transit spectra to form the template.    

Our final dataset consists of 31 red-channel and 31 blue-channel RVs.  The red-channel RVs have a median SNR of 37, resulting in a median RV error of 14\,m\,s$^{-1}$.  Meanwhile, the blue-channel RVs have a median SNR of 13 and a median RV error of 27\,m\,s$^{-1}$.  These RV error values are higher than those predicted by the MAROON-X Integration Time Calculator\footnote{https://www.gemini.edu/instrumentation/maroon-x/exposure-time-estimation}, which assumes a slowly rotating star.  However, these errors can be explained if the star is rotating rapidly, which would significantly reduce the RV precision \citep[for more detail,][provides a detailed discussion on the effect of rotational broadening on RV errors]{Bouchy2001}.

There are eight additional RVs on this system from the Tillinghast Reflecter Echelle Spectrograph (TRES), published in \cite{Irwin2018}.  The majority of the TRES exposures were not taken during the transit.  In addition, each individual TRES exposure is between 3600---3900s long, making them of limited use to characterizing the phase coverage of the (less than two hour long) transit.  However, the TRES RVs are useful for constraining the Keplerian motion of LP\,261-75C, which is important to characterize given the obvious slope in the MAROON-X data.

\begin{figure*}
    \centering
    \includegraphics[width=0.9\linewidth]{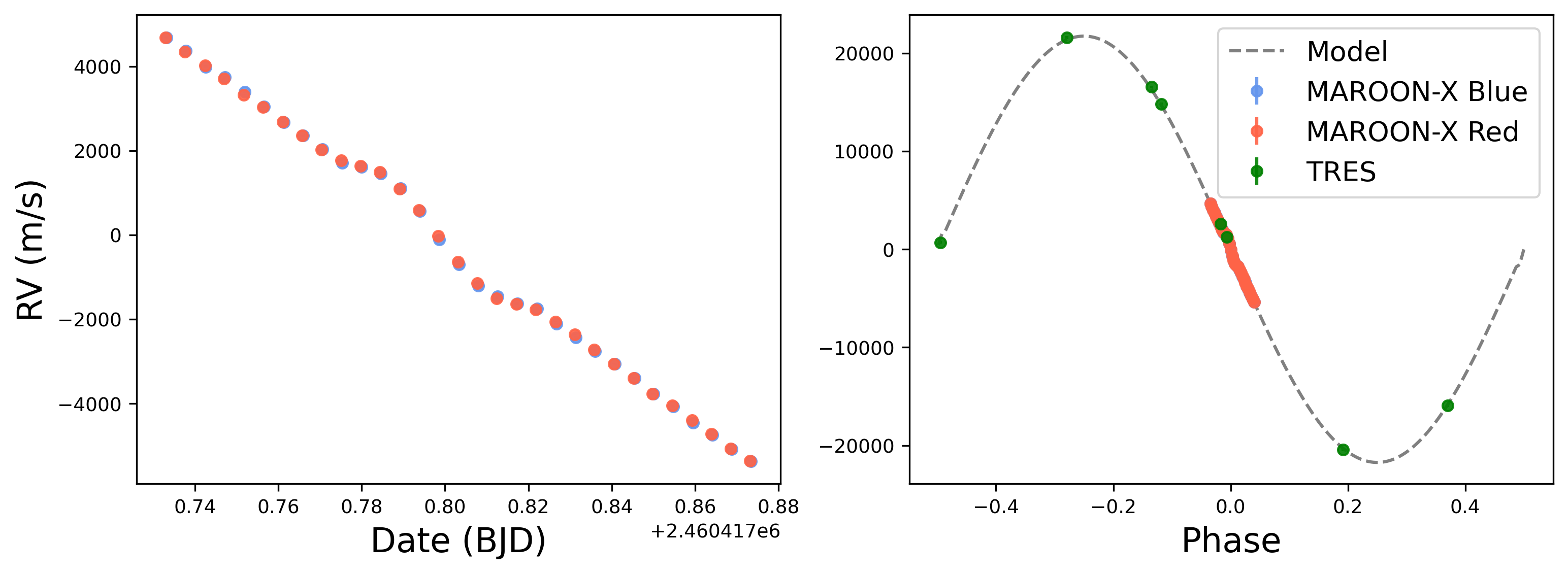}
    \caption{\textbf{Left:} The red- and blue-channel MAROON-X RVs of LP\,261-75A.  The RV uncertainties are too small to be visible in this plot.  \textbf{Right:} All of the available RV data on LP\,261-75A, with a fit model of the orbit of LP\,261-75C in gray.  The RV slope observed in the MAROON-X data is consistent with the brown dwarf's orbit inferred from the TRES data.}
    \label{fig:TOI1779_RV}
\end{figure*}

Figure~\ref{fig:TOI1779_RV} shows our newly-collected data alongside past RV data from TRES \citep{Irwin2018}, with an overplotted RV model for a 1.88d brown dwarf with a mass of 67\,$M_J$.  The dramatic slope in our data ($>5$\,km\,s$^{-1}$ over a few hours) is consistent with the known orbit of LP\,261-75C and is thus not concerning.  We provide a full table of the RVs collected with MAROON-X in Table~\ref{tab:RVs} in Appendix~\ref{appendix:rvs}.

\section{Analysis}
\label{sec:analysis}

\subsection{The Orbital Parameters of LP 261-75C}
\label{ssec:photometry_fitting}


Given the sensitivity of the RM curve to the transiting object's impact parameter and radius, we attempted to refine the brown dwarf's transit parameters using the available photometry.  We performed an analysis that included both the available \tess and \textit{MEarth} data in order to maximize our precision on the orbital parameters of the brown dwarf.

To perform a preliminary estimate of the brown dwarf's orbital period and transit time, we used the \texttt{transitleastsquares} \citep{Hippke19} code, which searches for transit-like signals in the light curve.  \texttt{transitleastsquares} recovered the transit at an orbital period of 1.8817d and a transit time of 2460417.794 (selected because it coincides with the collection times of our RV data).  These values agree roughly with the values quoted by \cite{Irwin2018}, which found an orbital period of $P\,=\,1.8817205\,\pm\,0.0000011$\,days and a transit time of $t_0\,=\,2458159.731511\,\pm\,0.000020$\,BJD (which corresponds with $t_0\,=\,2460417.796\,\pm\,0.001$\,BJD).

We used both sectors of the \tess data for this analysis, using the available 2-minute cadence data.  We also used the PDCSAP outputs from the SPOC pipeline \citep{Twicken2018} which includes detrended and dilution-corrected data.  To allow for the simultaneous fitting of both the \tess and MEarth data, we downloaded the MEarth data from \cite{Irwin2018} and converted it into relative flux units (instead of relative magnitude units).

We fit the transits using \texttt{juliet} \citep{Espinoza19}, which uses the transit modeling code \texttt{batman} \citep{batman} and the nested sampling code \texttt{dynesty} \citep{Speagle2020}.  We used the formalism from \cite{Espinoza18} to sample the brown dwarf's inclination and radius, fitting the parameters $r_1$ and $r_2$ instead of $b$ and $R_C/R_A$.  We modeled the stellar surface with quadratic limb-darkening parameters and fit for those as well, using the sampling formalism from \cite{Kipping13}, which fits $q_1$ and $q_2$ instead of $u_1$ and $u_2$.  We fixed the orbital eccentricity at zero in accordance with the RV results from \cite{Irwin2018}, which showed that LP\,261-75C has $e\,<\,0.007$.  


We chose to detrend the data with a simple Gaussian Process (GP) model in order to account for the obvious rotation signal present in the \tess photometric data. We used the approximate Matern GP kernel from \texttt{celerite} \citep{celerite}, and fit the GP parameters simultaneously with the brown dwarf transits, assuming broad priors.  

The priors, as well as the results to our photometric fit are shown in Table~\ref{tab:fit_transit_planet}.  Our measurements for the period, transit time, inclination, and $R_C$ are listed in Table~\ref{tab:fit_transit_planet} and are in agreement with the values from \cite{Irwin2018}, meaning that the \tess data and MEarth data appear to be in agreement with one another.  A fit without GP detrending produced similar results, but with a lower Bayesian evidence and slightly less precise constraints on the orbital parameters.

We also used the photometry fits to measure the star's radius.  Assuming a circular orbit \citep[reasonable considering the RV analysis of]{Irwin2018}, the density of LP\,261-75A is related to the semi-major axis of C by the following equation \citep[see, e.g.,][for more details]{Winn2010a}:

\begin{equation}
    \rho_A + \bigg( \frac{R_C}{R_A} \bigg)^3 \rho_C = \frac{3 \pi}{G P^2} \bigg( \frac{a_C}{R_A} \bigg)^3
\end{equation}

We can then solve for $R_A$, finding

\begin{equation}
    R_A =\bigg( \frac{(M_A+M_C) G P^2}{4 \pi^2} \bigg)^{1/3} \bigg( \frac{a_C}{R_A} \bigg)^{-1} 
\end{equation}

Using the masses of LP\,261-75A and C from \cite{Irwin2018}, we find that $R_A\,=\,0.308\,\pm\,0.005$\,R$_\odot$.  This value is in close agreement with the $R_A\,=\,0.306\,\pm\,0.009$\,R$_\odot$ value from the TESS Input Catalogue Version 8.2 \citep[TIC][]{Paegert21}, which was derived using the observed radii of M dwarfs from \cite{Mann15}. 

\subsection{Spectral Rotational Broadening}
\label{ssec:broadening}

The amplitude of a transiting object's RM signal is related to both the system orientation and the projected rotational velocity ($v_{eq}$\,sin$i_\star$) of the host star \citep[see, e.g., ][]{Winn2010a}.  This degeneracy can decrease the sensitivity of the RM curve to the sky-projected obliquity value if the rotational velocity (either from the $v_{eq}$\,sin\,$i$ or the stellar rotation period) is poorly constrained. 

\begin{figure*}
    \centering
    \includegraphics[width=0.9\linewidth]{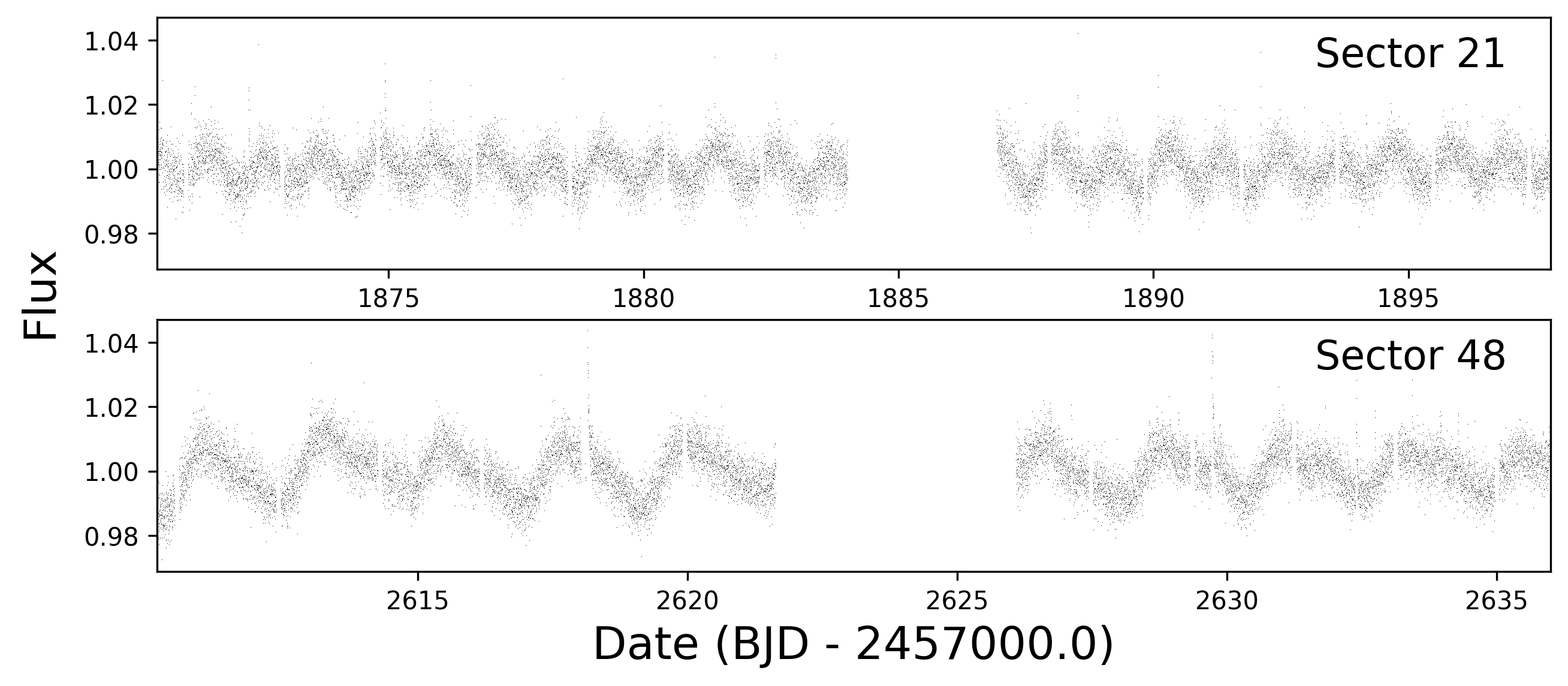}
    \caption{\tess 2-minute PDCSAP photometry of the LP\,261-75 system.  Outliers due to flares and transit events are removed in order to focus on the rotational modulation of the host star.}
    \label{fig:TOI1779_phot_rot}
\end{figure*}

There are two sectors' worth of \tess photometry for LP\,261-75A, shown in Figure~\ref{fig:TOI1779_phot_rot}.  Both sectors show signs of rotational modulation, with Sector 21 possessing a $\approx$1d signal and Sector 48 possessing a $\approx$2d signal.  \cite{Bowler2023} identifies this discrepancy as being due to a heavily spotted stellar surface, and identifies the longer 2.2d signal as being the true stellar rotation period. 

We can use this estimate of the stellar rotation period, along with the stellar radius from Section~\ref{ssec:photometry_fitting}, to measure the equatorial rotation velocity of the star.  If the star has a period of 2.2d, the expected equatorial rotation velocity is $7.04\,\pm\,0.17$\,km\,s$^{-1}$.  As the RM effect takes the \textit{projected} velocity into consideration, this value represents an upper limit on the rotation velocities we can expect to see.

We can independently measure the $v_{eq}$\,sin$i_\star$ of the star by measuring the rotational broadening present in the MAROON-X spectrum.  To do so, we performed an analysis similar to that done for TRAPPIST-1 in \cite{Brady23}, which was inspired by \cite{Gray05}.  In summary, we measured the rotation broadening of the highest-SNR LP\,261-75 spectrum by measuring the width of the cross-correlation function (CCF) of the LP\,261-75A spectrum with a MAROON-X spectrum of a known slow rotator with a similar spectral type.  We then artificially broadened the calibrator's spectrum and compared the width of its CCF the the CCF with the LP\,261-75A spectrum.  In this case, we use Barnard's Star as a calibrator, given both its slow rotation period \citep[145\,$\pm$\,15\,days,][]{Terrien22} and its spectral type of M4 \citep{Kirkpatrick1991}, which is very similar to the M4.5 star LP\,261-75.  

We measured the CCF of the LP\,261-75 spectrum with our Barnard's Star spectrum on an order-by-order basis.  After excluding all orders in which the CCF was not a single-peaked gaussian or otherwise seemed strongly affected by systematics, we estimated the $v_{eq}$\,sin$i_\star$ for each order and report the weighted mean and standard deviation of the $v_{eq}$\,sin$i_\star$ values as our final value.  We found that LP\,261-75 has line broadening consistent with a $v_\mathrm{eq}$sin$i$ of 7.78\,$\pm$\,0.48\,km\,s$^{-1}$.  

This value is slightly higher than the value expected from our photometrically-derived radius.  One possible explanation is that we have underestimated the masses of LP\,261-75A and C when calculating $R_A$, though this is unlikely to have a dramatic effect given the relatively weak dependence of $\rho_\star$ on mass. It is also possible that a systematic issue in our methodology is responsible for our measurement of $v_{eq}$\,sin$i_\star$.  Our calibrator star, Barnard's Star, is an old, slow-rotating field star, while LP\,261-75A is quite young and active, with emissive features in its spectrum.  Systematic differences between the two spectra due to their dramatic differences in age and activity features could result in broader, noisier CCFs when they are cross-correlated with one another, causing a higher measured $v_{eq}$\,sin$i_\star$.  These systematic errors may not necessarily be captured by our method.  However, the disagreement between the photometric radius and the measured $v_{eq}$\,sin$i_\star$ is not significant enough to be of serious concern.


As an additional check, we compared our results to those from \cite{Irwin2018}.  They performed a rotational broadening analysis of the TRES spectra of LP\,261-75 and recovered a rotation velocity of roughly $7.57 \pm 0.10$\,km\,s$^{-1}$, which is in agreement with our measured value.  However, their measured rotational broadening value is only slightly higher than the spectral resolution of the instrument, meaning that the reliability of the measurement was questionable.  As MAROON-X has a higher wavelength resolution than the TRES observations (85,000 vs 44,000), our measurement is well above our resolution and thus more reliable. However, due to the reasons discussed above, we believe it may be an overestimate.  We thus do not use it as a prior for our RM fits in Section~\ref{ssec:rm}. 

\subsection{Rossiter-McLaughlin Effect}
\label{ssec:rm}

\subsubsection{RM Model Comparison}
\label{sssec:model_comp}

As noted in \cite{Brown2017}, different models of the RM effect typically have different results for the final calculated $v_{eq}$\,sin$i_\star$, though the measurement of $\lambda$ tends to be consistent for aligned or near-aligned systems.  They specifically compared the results for the RM modeling method from \cite{Hirano2011}, developed for iodine cell spectrographs, with the results from \cite{Boue2013}, developed for CCF-based spectrographs.  However, \cite{Brown2017} found that the \cite{Boue2013} method tended to underestimate the $v_{eq}$\,sin$i_\star$ value, even for HARPS (which also lacks an iodine cell).

We thus fit several different RM model formulations to our RV data in order to evaluate the sensitivity of our results to the chosen model.  Firstly, we adopted the model from \cite{Hirano2011}.  This method is appropriate for \texttt{serval} data, as \texttt{serval} derives spectral RVs similarly to the iodine-cell method by finding the minimum in a $\chi^2$ valley of a template compared to the spectra. Additionally, the \cite{Hirano2011} model is the most complete RM model description---accounting for spectrograph resolution effects and other broadening effects which can impact both the $v\sin i$ and $\lambda$ values.  We then used \texttt{rmfit} \citep{Stefansson2022}, a publicly-available RM fitting code which implements the RM formulation described in \cite{Hirano2010}, which is similar to the \cite{Hirano2011} formulation but has a slightly less realistic equation for the stellar line profiles.  Next, we implemented a code that utilizes the RM anomaly equation from \cite{Ohta2005}, which is similar to the \cite{Hirano2011} model but does not include additional treatments due to, e.g., macroturbulent broadening from the star and the finite instrument resolution.  We finally fit the data using \texttt{starry} \citep{Luger19, starry}, a publicly-available code which can analytically model a stellar surface with arbitrary degrees of limb-darkening and then outputs the expected RV and transit signal from a user-specified transiting object.  

For our \cite{Hirano2011}, \cite{Ohta2005}, and \texttt{starry} model implementations, we estimated the fractional amount of light occulted $f$ numerically by modeling the star as a grid of points with quadratic limb-darkening.  We avoided using computationally faster analytic techniques due to the fact that $R_C/R_A$ is very large, invalidating many of the commonly-used approximations made in, e.g., \cite{Ohta2005}.

\begin{figure}
    \centering
    \includegraphics[width=0.9\linewidth]{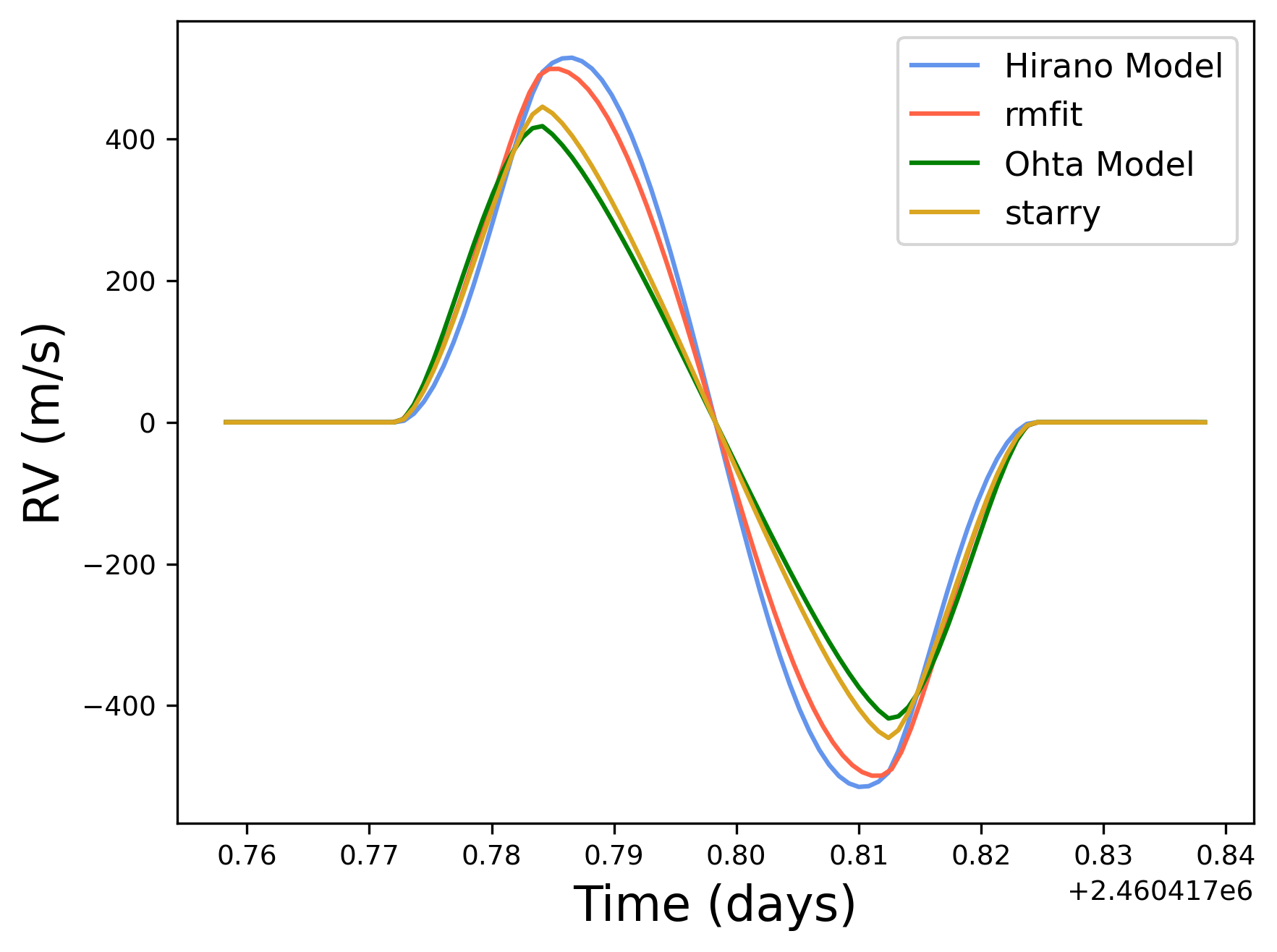}
    \caption{A comparison between the model outputs of the \cite{Hirano2011}, \texttt{rmfit}, \cite{Ohta2005}, and \texttt{starry} RM models.  For these models, we assume $\lambda\,=\,0^o$ $v_{eq}$\,sin$i\,=\,7.0$\,km\,s$^{-1}$, $u_1\,=\,0.4$, and $u_2\,=\,0.2$.  Both the \cite{Hirano2011} and \texttt{rmfit} models have $\beta\,=\,3.5$\,km\,s$^{-1}$ and $\zeta\,=\,1$\,km\,s$^{-1}$.  Finally, the \cite{Hirano2011} model assumes Gaussian line profiles, with $\gamma\,=\,0$, in order to allow for easier comparison to the \texttt{rmfit} models, which assume this formalism.  The addition of macroturbulent and instrumental broadening result in a larger RM amplitude for a given $v_{eq}$\,sin$i_\star$ value.  It is thus obvious that the \cite{Ohta2005} and \texttt{starry} models tend to under-predict the RM amplitude.}
    \label{fig:model_comp}
\end{figure}

Figure~\ref{fig:model_comp} compares these four models in several different cases, given an aligned planet, a rotation velocity of $v_{eq}$\,sin$i_\star\,=\,7.0$\,km\,s$^{-1}$, $u_1\,=\,0.4$, and $u_2\,=\,0.2$, which are close to the expected values of these parameters for this system.  It is obvious that the additional broadening considered in the \cite{Hirano2011} and \texttt{rmfit} model results in a higher RM amplitude, as well as a different shape in the RM curve. It is also clear that the \cite{Hirano2011} and \texttt{rmfit} models produce very similar results, demonstrating that the computationally simpler \cite{Hirano2010} model used in \texttt{rmfit} is a reasonable approximation for the \cite{Hirano2011} model.

We now describe the priors for our fits with the four models in detail.  The values for the priors are listed in Table~\ref{tab:fit_rm_planet}. For all four models, we performed our fits using \texttt{emcee} \citep{emcee}.  

We modeled the host star with quadratic limb-darkening, and allowed the limb-darkening coefficients $u_1$ and $u_2$ to vary uniformly.  We also included the constraint that the limb-darkening intensity must be positive on the stellar disk and decrease monotonically towards the edge of the disk.  For all models except \texttt{rmfit} (for which such a functionality was not available), we fit separate sets of limb-darkening coefficients for the red and blue channels of MAROON-X, as they encompass different wavelength ranges.  We also fit separate limb-darkening coefficients for the TRES data, though their exposures were likely too long and low-SNR to derive useful limb-darkening values.  We used \texttt{emcee} to perform MCMC fits of the parameters.

We adopted priors for the cosine of the inclination (cos\,$i$), the radius ratio ($R_C/R_A$) the period ($P$), and the transit time ($t_0$) of the brown dwarf from the fits to the transit data in Section~\ref{ssec:photometry_fitting}.  We implemented broad uniform priors (with a minimum value of 4\,km\,s$^{-1}$ and a maximum value of 12\,km\,s$^{-1}$) on the stellar rotational velocity ($v_{eq}$sin$i$) based roughly on our knowledge of the system's rotation period.  We included such broad priors in order to capture any systematic inaccuracies in $v_{eq}$sin$i$ by any of the tested models.

In order to accurately model the RV curve of several different instruments, we also included fits to the mean RV and jitter of each dataset ($\mu_{RV}$ and $\sigma_{RV}$), as well as the RV semi-amplitude of LP\,261-75C ($K$).  Our priors for these values (and all of the other parameters) are informed by the degree of scatter in the observed RV curves and are listed in the second column of Table~\ref{tab:fit_rm_planet}.

The \cite{Hirano2011} model includes several additional stellar parameters in addition to its rotation velocity, including the macroturbulence dispersion ($\zeta$), the gaussian velocity parameter ($\gamma$), and the the thermal velocity parameter ($\beta$).  We knew none of these parameters \textit{a priori} and thus included them in our fits.  As discussed in \cite{Hirano2011}, $\beta$ is related to the instrumental profile of the instrument, and we thus include a prior based on the R\,$\approx$\,85,000 resolution of MAROON-X, with $\beta\,\approx\,3.5$\,km\,s$^{-1}$.  We included broad normal priors on $\beta$ of N(3.5, 1.0)\,km\,s$^{-1}$ to account for additional influences of, e.g., microturbulence.  We set uniform priors on $\gamma$, constraining it such that $\gamma\,<\,1.5$\,km\,s$^{-1}$, similar to what was described in \cite{Hirano2011}.  We assumed a macroturbulent velocity of $\zeta$\,<\,1.5\,km\,s$^{-1}$, which is in-line with the low expected macroturbulent velocities of cool stars \citep{Soto2018}.  We included the same priors in the \texttt{rmfit} model, except for the fact that \texttt{rmfit} does not have a $\gamma$ parameter.

We simultaneously fit the RM signal with the Keplerian orbit of LP\,261-75C.  This is important because the Keplerian signal from the brown dwarf imposes a substantial near-linear trend on the RM curve, and accurately accounting for this background is important for constraining the RM parameters.  

The results for several key parameters for the different models are shown in Table~\ref{tab:rm_model_comp}.  We found that all four models had similar values of $\lambda$, but disagreed in terms of $v_{eq}$\,sin$i_\star$ and the limb-darkening parameters.  The models without the additional broadening effects had lower RV deviations in general, as well as less rounded RM curves.  Given the precise constraints on the $R_C/R_A$ and cos\,$i_C$, the models thus fit very high values of $v_{eq}$\,sin$i_\star$, and extreme values of $u_1$, and $u_2$ to compensate for these differences in shape and amplitude.  However, the fit value of $v_{eq}$\,sin$i_\star$ is far too high to be consistent with the known radius and rotation period from the star, and the limb-darkening parameters are highly discrepant from the ones expected for M dwarfs in \cite{Claret2012}.  This is in-line with the model expectations shown in \cite{Hirano2011}, which found that the \cite{Ohta2005} formulation tended to produce models with smaller RV amplitudes compared to other models for a fixed $v_{eq}$\,sin$i_\star$. The values calculated by the \cite{Hirano2011} model are less precise (likely due to the increased complexity of the fit), but the $v_{eq}$\,sin$i_\star$ ($7.2\,\pm\,0.4$\,km\,s$^{-1}$) is within 1$\sigma$ of the expected value given the star's rotation.  Additionally, the limb-darkening parameters are, while poorly-constrained, less discrepant from expectations.  The \texttt{rmfit} model recovers very similar results as the \cite{Hirano2011} model.

\begin{table*}
    \centering
    \begin{tabular}{|c|c|c|c|c|}
    \hline
    \textbf{Parameter} & \cite{Hirano2011}          & \texttt{rmfit}         & \cite{Ohta2005}        & \texttt{starry}        \\
    \hline
    $\lambda$ (degrees)& $4.8^{+15.3}_{-13.2}$      & $1.6^{+9.5}_{8.0}$     & $-0.4^{+9.2}_{-9.9}$  & $-0.7^{+7.5}_{-8.3}$   \\ 
$v_{eq}$\,sin$i_\star$ (km\,s$^{-1}$)&$7.1^{+0.3}_{-0.2}$& $7.0\pm 0.4$      & $8.5 \pm 0.3$         & $8.3^{+0.3}_{-0.2}$    \\
    $u_{1, blue}$      & $0.81^{+0.26}_{-0.36}$     &                        & $1.77^{+0.10}_{-0.13}$ & $1.58^{+0.12}_{-0.17}$ \\
    $u_{2, blue}$      & $-0.13^{+0.49}_{-0.30}$    &                        & $-0.83^{+0.13}_{-0.08}$& $-0.68^{+0.20}_{-0.13}$\\
    $u_{1, red}$       & $0.43^{+0.24}_{-0.33}$     &                        & $1.51^{+0.09}_{-0.12}$ & $1.98^{+0.24}_{-0.25}$ \\
    $u_{2, red}$       & $0.00^{+0.44}_{-0.24}$     &                        & $-0.66^{+0.17}_{-0.11}$& $-1.60^{+0.39}_{-0.37}$\\
    $u_{1, TRES}$      & $-1.71^{+1.35}_{-0.84}$    & -                      & $-1.67^{+1.40}_{-0.89}$& $-1.15^{+2.07}_{-1.33}$\\
    $u_{2, TRES}$      & $1.51^{+0.87}_{-0.91}$     & -                      & $1.51^{+0.89}_{-0.95}$ & $-1.22^{+2.44}_{-1.27}$\\
    $u_{2, all}$       & -                          & $0.99^{+0.44}_{-0.56}$ & -                      & -                      \\
    $u_{2, all}$       & -                          & $-0.51^{+0.79}_{-0.63}$& -                      & -                      \\
    \hline
    \end{tabular}
    \caption{A comparison between various fit parameters of interest for several different RM models.}
    \label{tab:rm_model_comp}
\end{table*}

Despite the differences in method, the four different models all recover a low $\lambda$ for the system, indicating that our measurement of the system's orientation is not model-dependent.  Given the more physically accurate parameters, we select the \cite{Hirano2011} model as our model moving forward.  

\subsubsection{Fitting The Obliquity}

We can now use the \cite{Hirano2011} model to find $\lambda$.  We can then calculate the true obliquity $\psi$ from the following equation:

\begin{equation}
    \mathrm{cos}(\psi) = \mathrm{cos}(i_\star) \mathrm{cos}(i_C) + \mathrm{sin}(i_\star) \mathrm{sin}(i_C) \mathrm{cos}(\lambda).
\end{equation}

As discussed in \cite{Masuda20}, we cannot accurately calculate the distribution of cos$(i_\star)$ by merely calculating cos$(i_\star)\,=\,\sqrt{1-(v\,\mathrm{sin}i_\star/v_{eq})^2}$. In order to account for the known correlations between $v_{eq}$, $v_{eq}$\,sin$i_\star$, $\lambda$, and $\psi$, we slightly adjusted our choice of fitting parameters to accurately reflect the known priors on $v_{eq}$.  Instead of fitting $v_{eq}$\,sin$i_\star$ directly, we instead fit $R_\star$, $P_{rot}$, and cos$i_\star$, with the following relation:

\begin{equation}
    v_{eq} \mathrm{sin}i_\star = \frac{2 \pi R_\star}{P_{rot}}  \sqrt{1-(\mathrm{cos}i_\star)^2}
\end{equation}

We thus directly fit for cos$i_\star$, allowing us to more accurately capture the value of $\psi$.  Our priors for $P_{rot}$ and $R_\star$ came from the stellar parameters listed in Table~\ref{tab:host} and the prior for cos$i_\star$ is uniform between -1 and 1.  All other priors are identical to those used in Section~\ref{sssec:model_comp} and are listed in the second column of Table~\ref{tab:fit_rm_planet}. We did not simultaneously fit the photometry, as our model is computationally intensive (and thus slow) to evaluate.  This choice is unlikely to affect the results given the very precise constraints from the photometry.

\begin{figure*}
    \centering
    \includegraphics[width=0.9\linewidth]{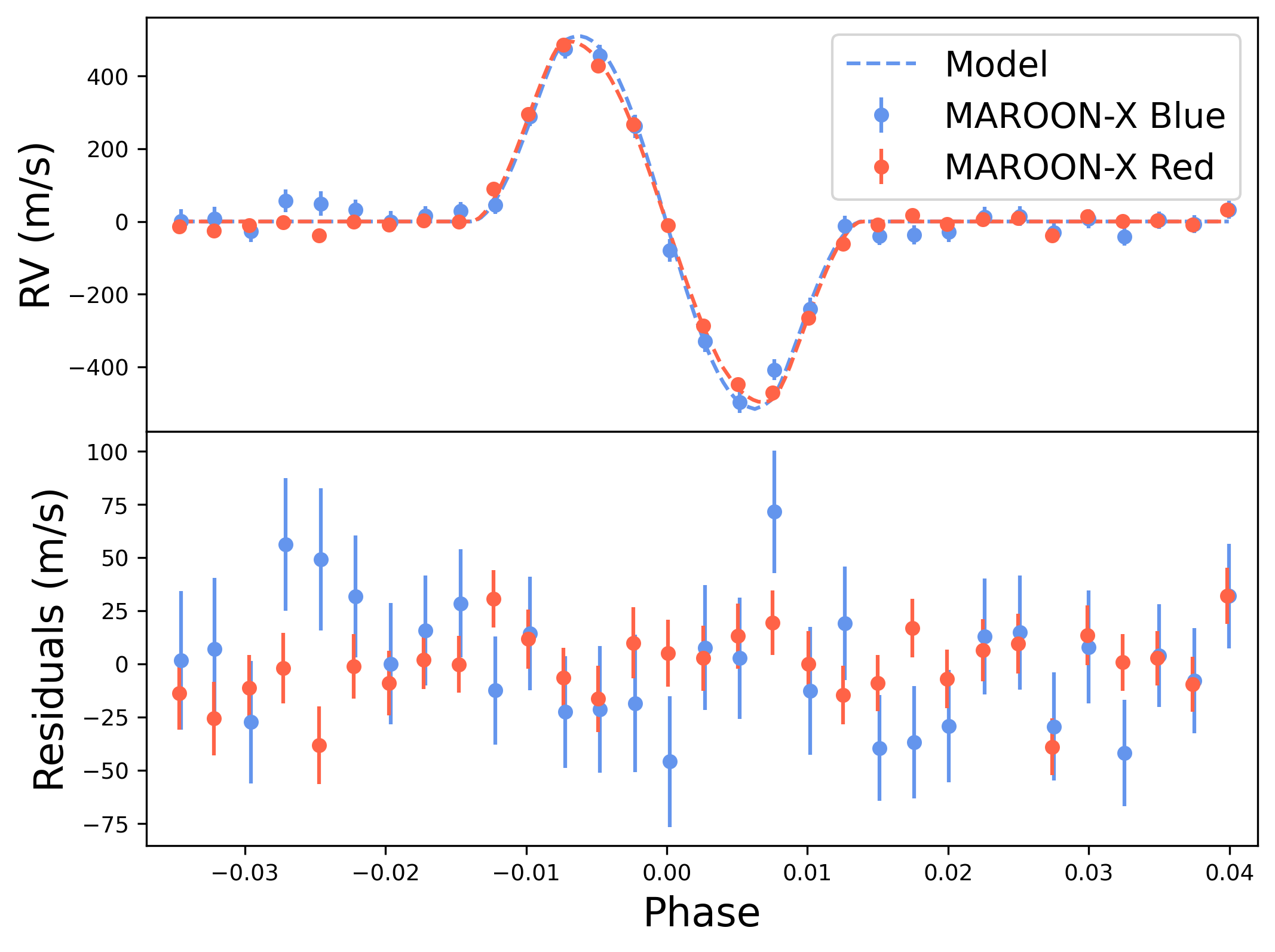}
    \caption{\textbf{Top:} Final RM fit to the MAROON-X RV data, with the Keplerian shifts subtracted out.  \textbf{Bottom:} Residuals of the RM fit to the MAROON-X data.}
    \label{fig:TOI1779_RM}
\end{figure*}

Our fit results for this model are listed in Table~\ref{tab:fit_rm_planet}.  Figure~\ref{fig:TOI1779_RM} shows the final fit RM model, with the Keplerian curve subtracted in order to emphasize the features of the RM signal. The model without the Keplerian subtracted is the gray line in the right panel of Figure~\ref{fig:TOI1779_RV}.  With the above formulae, we calculated that $v\,\mathrm{sin}i_\star\,=\,7.00^{+0.15}_{-0.16}$\,\kms.  We found that $\lambda\,=\,5^{+11}_{-10}$\,degrees and $\psi\,=\,14^{+8}_{-7}$\,degrees, meaning that the system is consistent with being aligned.  We also recovered a RV mass of $67.4\,\pm\,2.1\,$M$_J$ for LP\,261-75C, whose close agreement with the mass of $68.1\,\pm\,2.1\,$M$_J$ from \cite{Irwin2018} indicates that the RV slope in the MAROON-X data agrees with the RV signal from previous datasets.  Additionally, our fits merely recovered the priors for $\gamma$ and $\zeta$, indicating that those two parameters were not constrained by our data.

\section{Discussion}
\label{sec:discussion}

\subsection{Formation History}
\label{ssec:dynamics}

At its mass ($M_C\,>\,60\,$M$_J$), LP\,261-75C is likely to have formed via disk fragmentation, similar to how binary stars form \citep{Ma2014}.  It may have formed on a larger orbit and then migrated inwards.  Tidal interactions between LP\,261-75A and C could result in the angular momentum transfer from C’s orbit to the rotation of A, resulting in eventual spin-orbit synchronization.  Several other transiting M-dwarf brown-dwarf pairs, such as NGTS-7A \citep{Jackman2019} and TOI-263 \citep{Palle2021} have spin-orbit synchronization and could represent an end point for the evolution of this system.  Furthermore, the extremely short orbital period of a third system, ZTF J2020+5033, indicates that magnetic braking may be an important mechanism for shortening the orbits of brown dwarfs around fully-convective stars \citep{El-Badry2023}, even though previous studies \citep[such as][]{Schreiber2010} theorized that this mechanism might be dramatically weakened in such stars.  

The fact that LP\,261-75A currently has a longer rotational period (2.2d) than LP\,261-75C’s orbital period (1.88d) indicates that this system is not yet in its final orbital configuration.  If we assume that the tidal forces that guide this decay act on similar timescales to the tidal forces that would align LP\,261-75C, it is thus likely that the M dwarf has not had sufficient time to fully align its companion’s orbit.  We can also consider the obliquity damping timescale of stars with convective envelopes from \cite{Zahn1977} and \cite{Albrecht2012}:

\begin{equation}
    \tau_{CE} = 10^{10}\,\mathrm{yr} \times \bigg(\frac{M_C}{M_A}\bigg)^{-2} \bigg( \frac{a/R_A}{40} \bigg)^6.
\end{equation}

For the LP\,261-75 system, this timescale is around $\tau_{CE}\,\approx\,5*10^{8}$\,years,which is larger than the age of the system, though by less than an order of magnitude.  Given the imprecise nature of this relationship \citep[which was calibrated based off of empirical trends in observations in][]{Albrecht2012}, this is only weak evidence that the star could not have aligned the system.  Both the tidal decay and obliquity damping timescales seem to imply that the observed alignment ($\lambda\,=\,5^{+11}_{-10}$\,degrees, $\psi\,=\,14^{+8}_{-7}$\,degrees) is primordial in nature.  This is somewhat discrepant with \cite{Bowler2023}, which found that long-period brown dwarfs have a tendency to be less aligned with their host stars than long-period planets.  As longer-period brown dwarfs are less likely to tidally influence their host stars, that study implies that brown dwarfs do not tend to have highly aligned primordial obliquities.  

\begin{figure}
    \centering
    \includegraphics[width=0.9\linewidth]{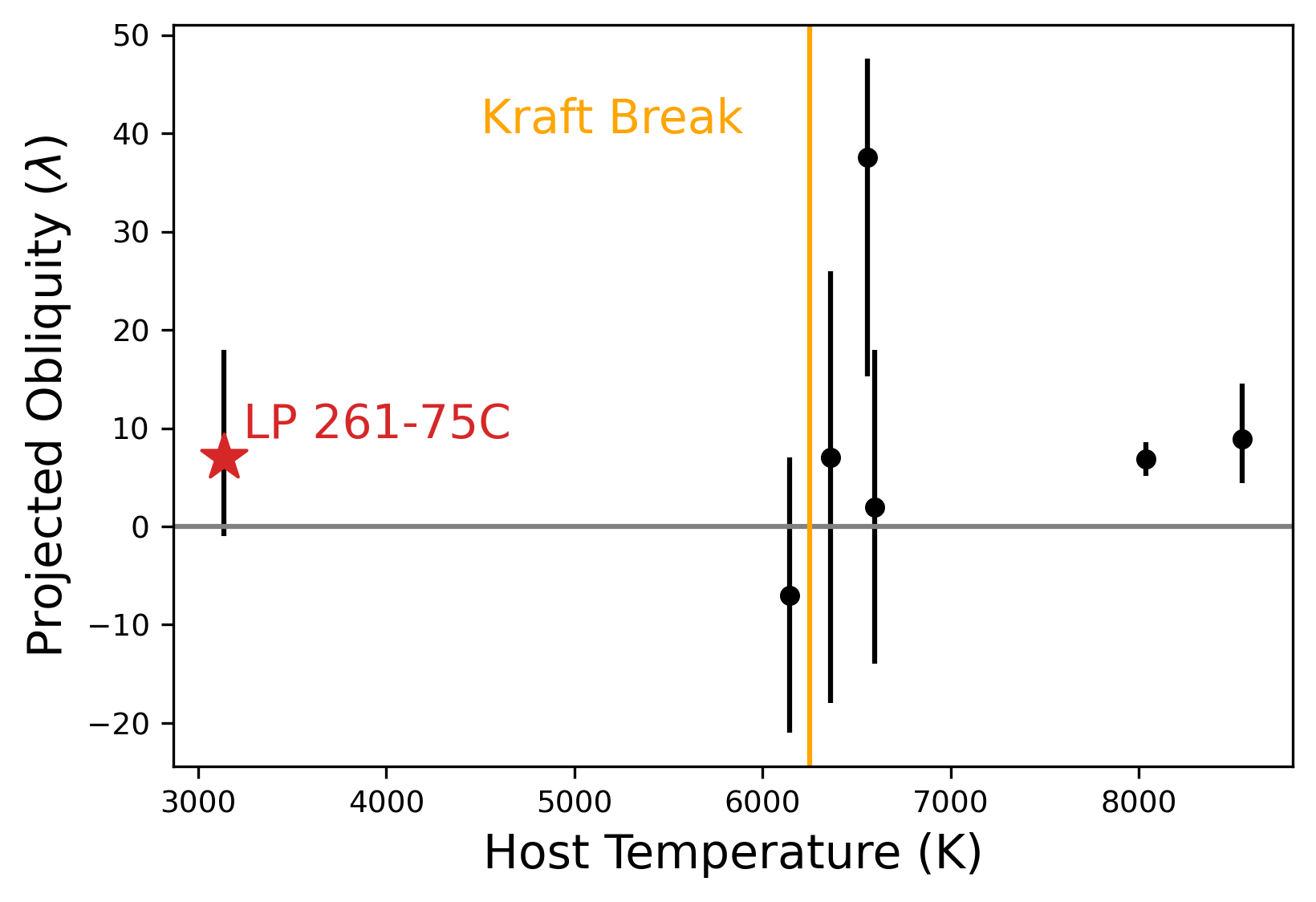}
    \caption{The host star $T_{eff}$ vs $\lambda$ for all systems with transiting brown dwarfs with measured obliquities.  The Kraft break is shown as an orange line, and the gray line corresponds to an aligned $\lambda\,=\,0^o$.  LP\,261-75C is highlighted with a red star.  The stellar temperatures are drawn from the \textit{TESS} Input Catalog \citep{Stassun19} and the obliquities are from \cite{Triaud2009}, \cite{Siverd2012}, \cite{Triaud2013}, \cite{Zhou2019}, \cite{Giacalone2024}, and \cite{Ferreira2024}.}
    \label{fig:BD_comp}
\end{figure}

However, as shown in Figure~\ref{fig:BD_comp}, which shows all of the obliquity measurements for brown dwarfs with RM measurements, it appears that the majority of short-period brown dwarfs have aligned (or nearly-aligned) orbits. This is true even for brown dwarfs orbiting host stars with temperatures above the Kraft break, which typically tend to host high-obliquity planets \citep{Winn2010b}.  The only exception to this trend is CoRoT-3b \citep{Triaud2009}, which has a relatively low-precision obliquity measurement.  

\cite{Albrecht2022} noted that most hot Jupiters with a companion-to-host mass ratio of $0.5 * 10^{-3}$ or higher tended to be aligned.  With a mass ratio of roughly $M_C/M_A\,=\,0.2$, LP\,261-75 falls above this cutoff.  Checking the literature, it appears that all of the brown dwarfs in Figure~\ref{fig:BD_comp} have masses above this cutoff, potentially explaining their alignment.  However, this is a purely empirical trend that is not necessarily supported by theory, especially for the planets orbiting stars hotter than the Kraft break.  We can estimate the obliquity damping timescales of these stars using Equation~3 from \cite{Albrecht2012}, which is based on the equilibrium tidal framework for radiative stars from \cite{Zahn1977} and calibrated based on observations of binary stars.  We find that the vast majority of brown dwarfs around hot stars are significantly younger than their damping timescales.  Despite this, they still appear to be aligned.  Even LP\,261-75, which is expected to have much more rapid obliquity damping due to its convective envelope, still has a timescale comparable to (or longer than) its age.  Given the growing sample of well-aligned brown dwarfs, it is starting to appear as though our current obliquity models do not accurately describe brown dwarfs.

LP\,261-75 thus joins the likes of AU Mic \citep{Addison2021} and K2-33 \citep{Hirano2024} as a young, aligned system around a fully convective M dwarf, as well as GPX-1b \citep{Benni2021, Ferreira2024} as a well-aligned young brown dwarf.  It will be necessary to study additional brown dwarfs around a variety of host stars in order to further explore the dynamical influences that they exert on their host stars.

\subsection{Comparisons to Isochrones}

One major problem that complicates the study of brown dwarfs is the degeneracy between their radius, luminosity, and age.  Unlike stars, a brown dwarf tends to cool, contract, and dim over the course of its lifetime, making it very difficult to estimate its parameters without evolutionary models.  As we can use our RV and photometric fits to estimate the precise mass and radius of LP\,261-75C (see Table~\ref{tab:bd}), we can compare its properties to an isochrone in order to evaluate the quality of brown dwarf models.  

\begin{figure}
    \centering
    \includegraphics[width=0.9\linewidth]{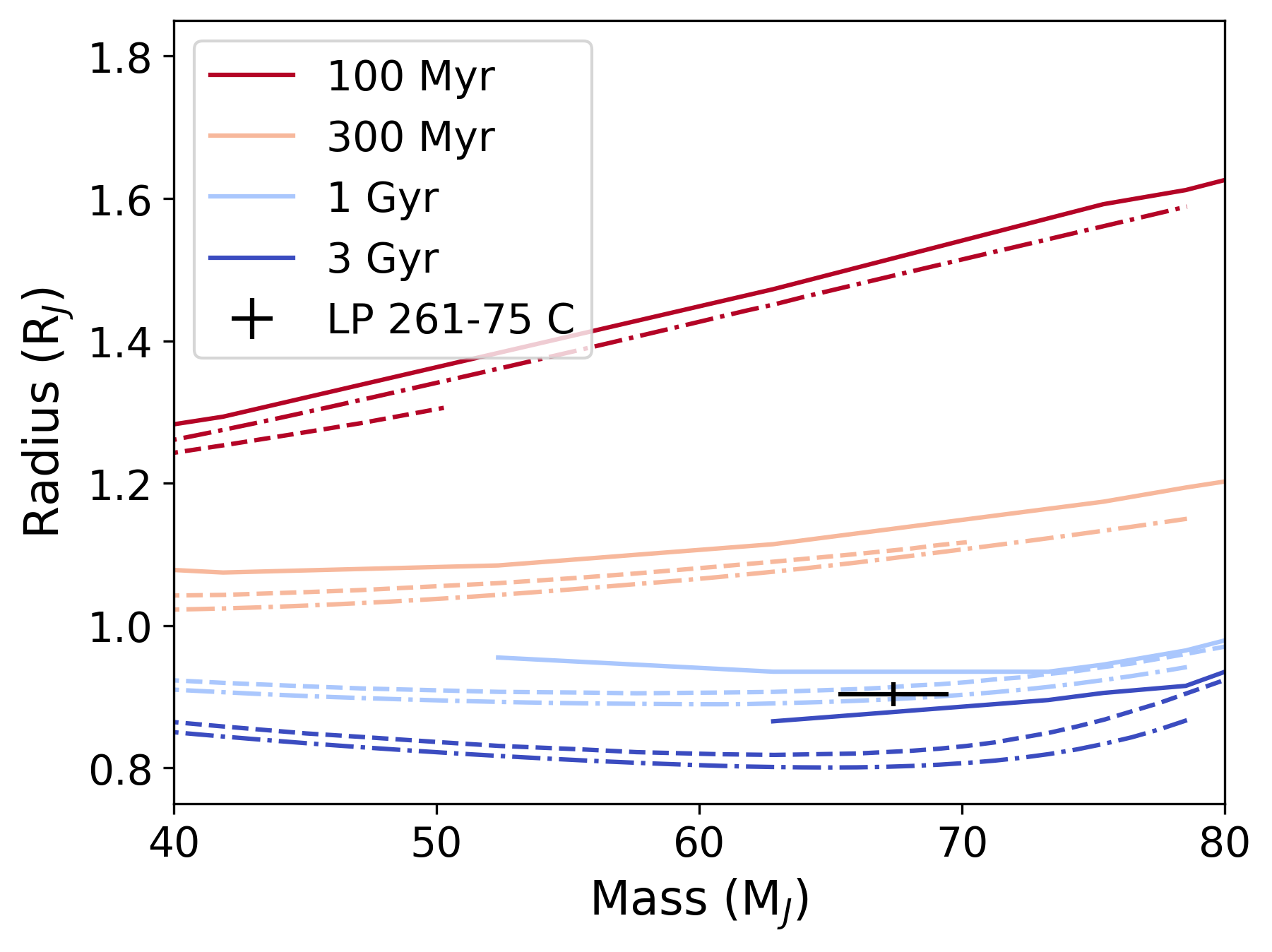}
    \caption{The mass and radius of LP\,261-75C, compared to brown dwarf isochrones from \cite{Baraffe15} (solid line), equilibrium chemistry \texttt{ATMO 2020} models \citep[dot-dashed line,]{Phillips2020}, and solar metallicity Sonora Bobcat models \citep[dashed line]{Marley2021}.  LP\,261-75C has a radius consistent with a much older system than the $\approx$100\,Myr-old LP\,261-75.}
    \label{fig:BD_models}
\end{figure}

Figure~\ref{fig:BD_models} shows brown dwarf isochrones from \cite{Baraffe15}, \cite{Phillips2020}, and \cite{Marley2021}, with LP\,261-75C overplotted.  LP\,261-75C appears to be far more compact than would be expected given the system's presence in the $\approx\,100$\,Myr old AB Doradus Moving Group.  This observation is consistent for all three models.  This is in stark contrast to the trend observed in other transiting brown dwarfs and hot Jupiters, which tend to have inflated radii compared to model expectations, which is likely due to irradiation from the host star \citep[see, e.g.,][]{Thorngren2018}.

This trend seems to imply that LP\,261-75C is far \textit{smaller} than we expect, which is strange given the formation history implied above.  The system has yet to enter full spin-orbit alignment, even though several other M dwarf-brown dwarf systems with similar (or younger) ages than the AB Doradus Moving Group are aligned.  In addition, the rapid rotation rate and the spectral emission features of LP\,261-75C imply an age far younger than that of a typical field star.  If we accept that the system is indeed young, then there must be something else responsible for the small radius of LP\,261-75C.  The brown dwarf's unusual traits point towards a gap in our understanding of the processes that control brown dwarf radii, and are worthy of future study.

\section{Summary and Conclusions}
\label{sec:conclusions}

LP\,261-75A is a nearby, fully-convective star with a transiting companion (LP\,261-75C) in the brown dwarf desert.  The system is in the $\approx\,100$\,Myr AB Doradus Moving Group, so it gives us an interesting opportunity to study the primordial obliquity of the companion of a fully-convective star.  

Our analysis of this system showcases the complexities involved in accurately measuring a star's rotation velocity $v_{eq}$\,sin$i_\star$.  After using the system's photometry to refine the radius of the host star, we found that our line broadening measurement recovered a $v_{eq}$\,sin$i_\star$ roughly consistent with expectations, though slightly higher than a physical value.  We concluded that the highly active nature of LP\,261-75A compared to our template star may have been responsible for the overestimate.  We also found that RM models that did not account for additional non-rotation broadening tended to overestimate $v_{eq}$\,sin$i_\star$.  This highlights the importance of using more complex models such as \cite{Hirano2011} in order to minimize biases.

By adopting the models from \cite{Hirano2011}, we were able to model and fit the RM effect present in the MAROON-X RVs on the system.  These models were able to recover a physically reasonable value of $v_{eq}$\,sin$i_\star$ and similar $\lambda$ values to other RM models, though the more complex \cite{Hirano2011} model had larger errors.  We found that the system has a projected obliquity of $5^{+11}_{-10}$\,degrees and a true obliquity of $14^{+8}_{-7}$\,degrees.  This large uncertainty in $\lambda$ despite our high-precision data is driven by the low ($b\,=\,0.08\,\pm\,0.05$) impact parameter of the system.  Within our errors, the orbit of LP\,261-75C appears to be aligned with the rotation axis of LP\,261-75A.  The system's youth and lack of spin-orbit coupling imply that the system is not old enough to have undergone full alignment, so this obliquity value may be primordial.  However, the orbits of brown dwarfs tend to be more aligned than expected given simple equilibrium tidal theory and comparisons to binary stars, so it is possible that we do not properly understand the timescales guiding obliquity evolution in brown dwarf systems.

We also took advantage of our new photometric and RV fits on the system to update the parameters of LP\,261-75C, finding them largely in agreement with those listed in \cite{Irwin2018}.  However, we also found, when comparing the brown dwarf's mass and radius to brown dwarf isochrones, that the LP\,261-75C appears to have a density more consistent with a brown dwarf ten times older than the LP\,261-75 system.  However, the age of LP\,261-75 is fairly well-established, given both its rapid rotation and moving group membership. Its unusual traits warrant future study.

\vskip 5.8mm plus 1mm minus 1mm
\vskip1sp

This research has also made use of NASA's Astrophysics Data System Bibliographic Services. 

The University of Chicago group acknowledges funding for the MAROON-X project from the David and Lucile Packard Foundation, the Heising-Simons Foundation, the Gordon and Betty Moore Foundation, the Gemini Observatory, the NSF (award number 2108465), and NASA (grant number 80NSSC22K0117). The Gemini observations are associated with program GN-2024A-FT-105 (PI: Brady).

Support for this work was provided by NASA through the NASA Hubble Fellowship grant \#HF2-51559 awarded by the Space Telescope Science Institute, which is operated by the Association of Universities for Research in Astronomy, Inc., for NASA, under contract NAS5-26555.


Some of the data presented in this paper were obtained from the Mikulski Archive for Space Telescopes (MAST) at the Space Telescope Science Institute. The specific observations analyzed can be accessed via \cite{TESS_Curves}.

\software{Astropy \citep{Astropy1, Astropy2, Astropy3}, batman \citep{batman}, dynesty \citep{Speagle2020}, emcee \citep{emcee}, juliet \citep{Espinoza19}, lightkurve \citep{lightkurve}, Numpy \citep{numpy}, PyAstronomy \citep{PyAstronomy}, rmfit \citep{Hirano2011, Stefansson2022}, Scipy \citep{2020SciPy-NMeth}, spectres \citep{Carnall17}, starry \citep{Luger19, starry}, tpfplotter \citep{Aller20}}

\facility {Exoplanet Archive}

\bibliography{manuscript}

\appendix
\setcounter{table}{0}
\renewcommand{\thetable}{A\arabic{table}}

\section{RVs}
\label{appendix:rvs}


\begin{table}[h]
\centering
\begin{tabular}{|c|c|}
\hline
Column Name & Description                                              \\ \hline
channel     & Channel of observation (``red'' or ``blue'')             \\
bjd         & Date of observation (BJD)                                \\
rv          & Radial velocity (m/s)                                    \\
erv         & Radial velocity error (m/s)                              \\
sn\_peak     & Peak SNR of spectrum                                     \\
exptime     & Exposure time (s)                                        \\
berv        & Barycentric radial velocity (m/s)                        \\
airmass     & Airmass of observation                                   \\
dLW         & Differential line width (1000 m$^2$/s$^2$)               \\
e\_dLW       & Differential line width error  (1000 m$^2$/s$^2$0        \\
crx         & Chromatic index (m/s/Np)                                 \\
e\_crx       & Chromatic index error (m/s/Np)                           \\
irt\_ind1    & Calcium infrared triplet (8500.4 \AA \,line) index       \\
irt\_ind1\_e  & Calcium infrared triplet (8500.4 \AA\, line) index error \\
irt\_ind2    & Calcium infrared triplet (8544.4 \AA\, line) index       \\
irt\_ind2\_e  & Calcium infrared triplet (8544.4 \AA\, line) index error \\
irt\_ind3    & Calcium infrared triplet (8664.5 \AA\, line) index       \\
irt\_ind3\_e  & Calcium infrared triplet (8664.5 \AA\, line) index error \\
halpha\_v    & H$\alpha$ index                                          \\
halpha\_e    & H$\alpha$ index Error                                    \\
nad1\_v      & Sodium doublet (5891.6 \AA\, line) index                 \\
nad1\_e      & Sodium doublet (5891.6 \AA\, line) index error           \\
nad2\_v      & Sodium doublet (5897.6 \AA\, line) index                 \\
nad2\_e      & Sodium doublet (5897.6 \AA\, line) index error          \\
\hline
\end{tabular}
\caption{A description of the information contained in our online dataset on LP\,261-75A.}
\label{tab:RVs}
\end{table}

\section{Photometry Fits}
\label{appendix:phot_fits}

\begin{table}[h]
    \centering
    \begin{tabular}{|c|c|c|c|c|}
    \hline
    \textbf{Parameter} & \textbf{Prior}         & \textbf{Target} & \tess & MEarth \\
    \hline
    $P_C$ (d)                 & N(1.88172, 0.00004)    & $1.88172236 \pm 0.00000009$&                        &                        \\
    $t_{0, C}$ (BJD)             & N(2460417.794, 0.1)    & $2460417.79832 \pm 0.00008$&                        &                        \\
    $r_1$                   & U(0, 1)                & $0.3848 \pm 0.0311$        &                        &                        \\
    $r_2$                   & U(0, 1)                & $0.2938 \pm 0.0022$        &                        &                        \\
    $\frac{a}{R_A}$         & U(10, 20)              & $14.89 \pm 0.10$           &                        &                        \\ 
    $e$                     & 0 (Fixed)              &                            &                        &                        \\
    $\omega$ (degrees)      & 90 (Fixed)             &                            &                        &                        \\
    $\mu_{Flux}$ (ppm)      & N(0, 10$^5$)           &                            & $-194 \pm 264$         & $554 \pm 1113$           \\
    $\sigma_{Flux}$ (ppm)   & ln U(10$^{-6}$, 10$^6$)&                            & $0 \pm 18$            & $6498 \pm 45$          \\
    $\rho_{GP}$             & N(2.23, 2)             & $0.104 \pm 0.007$          &                       &                        \\            
    $\sigma_{GP}$ (ppm)     & ln U(10$^{-6}$, 10$^6$)&                            & $3680 \pm 120$        & $12270 \pm 860$        \\ 
    $q_1$                   & U(0, 1)                &                            & $0.27 \pm 0.12$       & $0.33 \pm 0.13$ \\
    $q_2$                   & U(0, 1)                &                            & $0.45 \pm 0.15$       & $0.33 \pm 0.12$ \\
    \hline
    $R_C/R_A$               &                        & $0.2938 \pm 0.0022$        &                        &                        \\
    $b$                     &                        & $0.077 \pm 0.052$          &                        &                        \\
    cos\,$i$                &                        & $0.0052 \pm 0.0031$        &                        &                        \\
    $u_1$                   &                        &                            & $0.47 \pm 0.06$            & $0.38 \pm 0.07$ \\
    $u_2$                   &                        &                            & $0.05 \pm 0.16$            & $0.19 \pm 0.16$ \\
    \hline
    \end{tabular}
    \caption{The brown dwarf parameters from the \texttt{juliet} transit fit to the \tess and MEarth photometry.}
    \label{tab:fit_transit_planet}
\end{table}

\section{RV Fits}
\label{appendix:rm_fits}

\begin{table}[h]
    \centering
    \begin{tabular}{|c|c|c|c|c|c|}
    \hline
    \textbf{Parameter}         & \textbf{Prior}            & \textbf{Target} & \textbf{MX Blue} & \textbf{MX Red} & \textbf{TRES} \\
    \hline
    $P_C$ (d)                  & N(1.88172235, 0.00000009) & $1.88172235^{+0.00000009}_{-0.00000010}$ & & & \\
    $t_{0, C}$ (BJD)           & N(2460417.79829, 0.00009) & $2460417.79824^{+0.00008}_{-0.00007}$ & & &\\
    R$_C$/R$_A$                & N(0.2919, 0.0029)         & $0.2935^{+0.0027}_{-0.0026}$ & & &\\
    cos\,$i_C$                 & N(0.0061, 0.0035)         & $0.0029^{+0.0034}_{-0.0025}$ & & &\\
    $\lambda$ (degrees)        & U(-180, 180)              & $4.8^{+11.3}_{-10.2}$ & & &\\ 
    $K$ (km\,s$^{-1}$)         & N(21.780209, 10)          & $21.75 \pm 0.02 $ & & &\\
    $\zeta$ (km\,s$^{-1}$)     & U(0, 1.5)                 & $0.8 \pm 0.5$ & & & \\
    $\beta$ (km\,s$^{-1}$)     & N(3.5, 1)                 & $4.0^{+0.7}_{-0.8}$& & & \\
    $\gamma$ (km\,s$^{-1}$)    & U(0, 1.5)                 & $0.9^{+0.4}_{-0.6}$& & & \\
    $\mu_{RV}$ (km\,s$^{-1}$)  & U(-10, 10)                &  & $1.76 \pm 0.01$ & $1.55 \pm 0.01$ & $-5.29 \pm 0.09$\\
    $\sigma_{RV}$ (m\,s$^{-1}$)& U(0, 20000)               &  & $10^{+8}_{-7}$& $9^{+5}_{-5}$& $212^{+120}_{-94}$\\
    $u_1$                      & U(-3, 3) &                & $0.75^{+0.25}_{-0.35}$ & $0.40^{+0.24}_{-0.34}$ & $-1.71^{+1.37}_{-0.87}$\\
    $u_2$                      & U(-3, 3) &                & $-0.07^{+0.48}_{-0.31}$& $0.01^{+0.48}_{-0.24}$ & $1.48^{+0.83}_{-0.94}$ \\
    $R_\star$ (\Rsun)          & N(0.308, 0.005)           & $0.309\pm 0.005 $ & & &\\
    P$_{\mathrm{rot}}$ (d)     & N(2.214, 0.040)           & $2.214 ^{+0.037}_{-0.038} $ & & &\\
    cos$i_\star$               & U(-1, 1)                  & $-0.00^{+0.19}_{-0.20}$ & & &\\
    \hline
    \end{tabular}
    \caption{The brown dwarf parameters from the MCMC fit to the RV data.}
    \label{tab:fit_rm_planet}
\end{table}

\end{document}